\documentclass[twocolumn,amsmath,amssymb]{revtex4}

\usepackage{graphicx}
\usepackage{dcolumn}
\usepackage{bm}

\begin{document}

\title{Universal and non-universal properties of wave chaotic scattering systems}

\author{Jen-Hao Yeh$^{1}$}
\author{James A. Hart$^{2}$}
\author{Elliott Bradshaw$^{2}$}
\author{Thomas M. Antonsen$^{1,2}$}
\author{Edward Ott$^{1,2}$}
\author{Steven M. Anlage$^{1,2}$}

\affiliation{$^{1}$Electrical and Computer Engineering Department,
University of Maryland, College Park, MD 20742-3285 \\
$^{2}$Physics Department, University of Maryland, College Park, MD
20742-4111}

\date{\today}

\begin{abstract}
The application of random matrix theory to scattering requires
introduction of system-specific information. 
This paper shows that the average impedance matrix, which
characterizes such system-specific properties, can be
semiclassically calculated in terms of ray trajectories between
ports.
Theoretical predictions are compared with experimental results for a
microwave billiard, demonstrating that the theory successfully
uncovered universal statistics of wave-chaotic scattering systems. 
\end{abstract}
\maketitle

Wave systems appear in diverse branches of physics, such as quantum
mechanics, electromagnetics and acoustics. However, solving the wave
equations can be difficult, particularly in the short wavelength
limit \cite{gutzwiller_book}. 
Furthermore, even if exact solutions were feasible, there may be
uncertainties in the locations of boundaries or in parameters
specifying the system, and the desired wave quantities can be extremely sensitive to such uncertainties when the wavelength is short. 
Thus, rather than seeking solutions for specific systems, it is
often convenient to create statistical models which reproduce
generic properties of the system \cite{haake_book}. 
This is the motivation for the application of random matrix theory
to wave-chaotic systems, in which it is conjectured that useful
statistics can be obtained by replacing the exact Hamiltonian or
scattering matrices by random matrices drawn from an appropriate
ensemble. 
Here, by wave-chaotic we mean that, in the small wavelength limit,
the behavior of the wave system is described by ray orbit
trajectories that are chaotic \cite{S1}. 
Although such formulations cannot predict details of any particular
wave system, they do predict the distribution of properties in an
ensemble of related wave-chaotic systems. Random matrix theory is
also hypothesized to predict the statistical properties of a
\emph{single} wave-chaotic system evaluated at different frequencies
(in, e.g., the cases of acoustics or electromagnetics) or energies
(in the case of quantum mechanics).
See
Refs.~\cite{Beenakker_review_RMT,Alhassid_review,Quantum_graphs_review}
for reviews of the theory, history, and the wide range of
applications
of random matrix theory. 

In this paper, random matrix theory is applied to model the
scattering behavior of an ensemble of wave-chaotic systems coupled
to the outside world through a single scattering channel (the
generalization to larger numbers of scattering channels is
straightforward \cite{Hart}). Such scattering systems have been
studied extensively, with most work focusing on the scattering
parameter $S$
\cite{Dyson_original,Poisson_Kernel_Original,Doron_Smilansky_Poisson,Poisson_including_internal,
Poisson_including_internal_2},
which is the ratio between the reflected waves and the incident
waves in the channel. 
Here we consider ensembles of systems whose distribution of
scattering parameters are well-described by the so-called Poisson
kernel
\cite{Dyson_original,Poisson_Kernel_Original,Doron_Smilansky_Poisson,Kuhl_poisson_kernel}.
The Poisson kernel characterizes the probability density for
observing a particular scattering parameter $S$ in terms of the
average scattering parameter $\bar{S}$. It represents contributions
to the scattering behavior from elements of the system which are not
random, such as the prompt reflection from the interface between the
scattering channel and the chaotic system. In addition, rays within
the scattering region which return to the scattering channel without
ergodically exploring the chaotic dynamics also affect $\bar{S}$
\cite{Poisson_including_internal,Bulgakov_Gopar_Mello_Rotter,
Weidenmuller}. The ability to determine $\bar{S}$ from first
principles, thus incorporating all non-universal effects, would
dramatically improve our understanding and ability to uncover
universal fluctuations in a host of wave phenomena. Because
$\bar{S}$ is the only parameter in the Poisson kernel, methods for
finding it are of interest. Even though $\bar{S}$ can be estimated
from experimental ensemble data, \emph{predicting} it from first
principles has so far not been addressed in general (although it has
been done for the specific case of quantum graphs \cite{Kottos}). In
this paper we show how to semiclassically obtain $\bar{S}$, and we
experimentally verify the accuracy and utility of our technique.

A quantity equivalent to the scattering parameter $S$ is the
impedance, $Z=Z_{0}(1+S)/(1-S)$, where $Z_0$ is the characteristic
impedance of the scattering channel. Because non-universal
contributions manifest themselves in $Z$ as simple additive
corrections, we use $Z$ in much of our discussion
\cite{Lossy_impedance_Savin_Fyodorov,Henry_paper,henrythesis}. 
We note that impedance is a meaningful concept for all scattering
wave systems.
In linear electromagnetics, it is defined via Ohm's law as
$\hat{V}=Z\hat{I}$, where $\hat{V}$ represents the phasor voltage
difference across the attached transmission line (the system's port)
and $\hat{I}$ denotes the phasor current flowing through the
transmission line. In acoustics, the impedance is the ratio of the
sound pressure to the fluid velocity. A quantum-mechanical quantity
corresponding to impedance is the reaction matrix
\cite{Alhassid_review,Lossy_impedance_Savin_Fyodorov}. 
In what follows, our discussion will use language appropriate to
scattering from a microwave cavity excited by a small antenna fed by
a transmission line (the setting for our experiments).

With the transformation to impedance, if $S$ is distributed
according to the Poisson kernel, we find that the impedance can be
represented as \cite{Hart}
\begin{equation}\label{eq:chaotic_impedance_statistics}
    Z=i X_{avg}+R_{avg}i\xi,
\end{equation}
where in the lossless case $R_{avg}$ and $X_{avg}$ are the real and
imaginary parts of the impedance $Z_{avg}$ based on the average
scattering parameter, where $Z_{avg}\equiv
Z_{0}(1+\bar{S})/(1-\bar{S})$ and $i\xi$ (which we call the
normalized impedance) is a Lorentzian distributed random variable
with median 0 and width 1. With uniform loss (e.g., due to an
imaginary part of a homogeneous dielectric constant in a microwave
cavity), $R_{avg}$ and $X_{avg}$ are the analytic continuations of
the real and imaginary parts of $Z_{avg}$ as $k\rightarrow
k+ik/(2Q)$, where $Q\gg 1$ is the quality factor of the closed
system, and $k$ denotes the wavenumber of a plane wave. The
normalized impedance $i\xi$ of the lossy system has a universal
distribution which is dependent only on the ratio $k/(2Q \Delta k)$,
where $\Delta k$ is the mean spacing between modes
\cite{Lossy_impedance_Savin_Fyodorov,Henry_paper}.

We find that $Z_{avg}$ can be evaluated directly in the
semiclassical limit \cite{Hart} as
\begin{equation}\label{eq:Z_avg_introduction}
    Z_{avg}=Z_{R}+R_{R}\left\langle\sum_{b(l)}C_{b(l)}e^{i A_{b(l)}}\right\rangle,
\end{equation}
where $R_{R}$, the radiation resistance, and $X_{R}$, the radiation
reactance, are the real and imaginary parts of the radiation
impedance $Z_{R}$, which represents the impedance the system would
have if all the energy which coupled into the system was absorbed
before coupling back out \cite{Henry_paper,henrythesis},
$\langle\ldots\rangle$ indicates a suitable ensemble averaging (to
be discussed further), $b(l)$ is an index over all classical
trajectories which leave the port and return to the port location
with total path length $l$, $C_{b(l)}$ is a function of the
stability of the trajectory indexed by $b(l)$, and $A_{b(l)}$ is the
action along the trajectory $b(l)$ \cite{Hart}. $C_{b(l)}$ includes
the initial phase shift and a geometrical factor that takes account
of the spreading of the ray tube along its path. 
Here it has been assumed that the port radiates isotropically from a
location far from the two-dimensional cavity boundaries. These
parameters can all be determined from the geometry of the scatterer
and location of the port \cite{Hart}.

The purpose of this letter is to test the accuracy and usefulness of
Eq. (\ref{eq:Z_avg_introduction}). In practice, we take account of a
finite number of ray trajectories according to their length $l$.
Therefore, in Eq. (\ref{eq:Z_avg_introduction}) we replace the
summation by $\displaystyle\sum_{b(l),L}$ which signifies that the
sum is now over all trajectories $b(l)$ with lengths up to $L$,
$l\leq L$. 

\begin{figure}
\includegraphics[height=2.4in,width=3.0in]{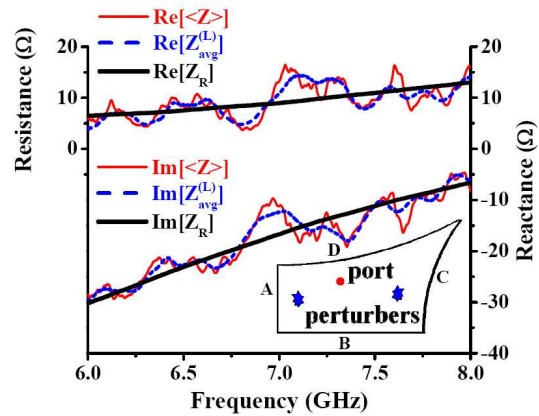}
\caption{(Color online) Plot of the impedance from the average of
100 cavity realizations $\langle Z\rangle$, versus frequency from 6
to 8 GHz with perturbers inside the cavity. Shown are the real
(three upper curves) and the imaginary parts (three lower curves) of
the impedance for the theory ($Z^{(L)}_{avg}$ with $L=200$ cm, blue
dashed) and the experiment ($\langle Z\rangle$, red solid), as well
as the measured radiation impedance of the port ($Z_{R}$, black
thick). Inset: The wave chaotic two-dimensional cavity with
perturbers and a single port.} \label{ensemble_impedance}
\end{figure}

In order to test our theoretical prediction, experimental tests are
carried out on a quasi-two-dimensional microwave cavity with a
single port \cite{S1, Sameer} (see Fig. \ref{ensemble_impedance},
inset), where the length of Wall A is 21.6 cm, the length of Wall B
is 43.2 cm, the distance of the port to the nearest wall (Wall D) is
7.5 cm, and the height of the cavity is 0.8 cm. For the frequency
explored (6 to 18 GHz), higher-order vertical modes are cutoff so
that the waves are quasi-two-dimensional. Furthermore, the
corresponding wavelengths (1.7 cm to 5.0 cm) can be regarded as in
the semiclassical regime, and the cavity shape yields chaotic ray
trajectories. 
We excite the cavity by means of a single coaxial probe whose
exposed inner conductor extends from the top plate and almost makes
electrical contact with the bottom plate of the cavity \cite{S1}.

The radiation impedance $Z_{R}$ in Eq. (\ref{eq:Z_avg_introduction})
is measured by covering the four side walls of the cavity with
microwave absorbers. Normalizing the measured impedance with the
radiation impedance has been used to remove the non-universal
properties due to the coupling of the port and the cavity \cite{S1,
Sirko}. Here we further consider the non-universal properties due to
short ray trajectories by adding the summation term in Eq.
(\ref{eq:Z_avg_introduction}).

To verify that Eq. (\ref{eq:Z_avg_introduction}) describes
non-universal characteristics of wave-chaotic systems, we first
proceed to determine universal statistics by applying the ensemble
average. Two irregular-shaped pieces of metal
(with the maximum diameters 7.9 cm and 9.5 cm) 
are added as perturbers in the wave-chaotic system that is shown in
the inset of Fig. \ref{ensemble_impedance}, where the circular dot
shows the port and the two starlike objects represent the
perturbers. The locations of the two perturbers inside the cavity
are systematically changed to produce a set of 100 realizations for
the ensemble \cite{Sameer}.
Typically, the shifts of resonances between two realizations are
about one mean level spacing. 
The scattering parameter $S$ is measured from 6 to 18 GHz, covering
roughly 1070 modes of the cavity.
After the ensemble average, longer ray trajectories have higher
probability of being blocked by the two perturbers, and therefore,
the main non-universal contributions are due to short ray trajectories. 
We compare the measured ensemble averaged impedance $\langle
Z\rangle$ and the theoretical impedance $Z_{avg}^{(L)}$ that is
calculated from Eq. (\ref{eq:Z_avg_introduction}) with the sum up to
the maximum trajectory length $L=200$ cm, corresponding to a total
of 584 trajectories. 
We multiply each term in the sum, Eq. (\ref{eq:Z_avg_introduction}),
by a weight equal to the fraction of perturbation configurations
that are not intercepted by the trajectory corresponding to that
term. The result is shown in Fig. \ref{ensemble_impedance} where the
three upper curves are the real part of the impedance (resistance)
and the three lower curves are the imaginary part (reactance). The
experiment curves (red solid) follow the trend of the radiation
impedance curves (black thick), and the theory curves (blue dashed)
reproduce most of the fluctuations in the experiment curves with
only a modest number of trajectories. The good agreement between the
measured data and the theoretical prediction verifies that the new
theory, Eq. (\ref{eq:Z_avg_introduction}), predicts the
non-universal features embodied in the ensemble averaged impedance.
We believe that the differences between the measured data and the
theory are due to the finite number of realizations in the ensemble
and diffraction
effects that are not taken into account in the theory. 

\begin{figure}
\includegraphics[height=2.4in,width=3.0in]{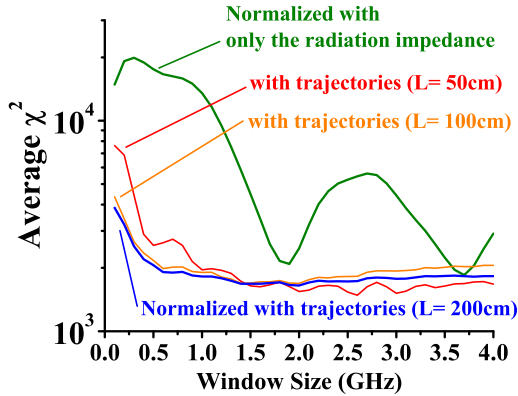}
\caption{(Color online) The average $\chi^{2}$ of distributions of
the phase of the scattering parameter $\varphi_{s}$ on a semi-log
scale, where the scattering parameters are calculated from impedance
normalized with only the radiation impedance (green) and with ray
trajectories according to the maximum trajectory length from $L=50$
cm (red) up to $L=200$ cm (blue), versus frequency window sizes from
0.1 to 4.0 GHz.} \label{chi_square}
\end{figure}

We further test our theory by consideration of the statistics of the
scattering parameter for an ensemble of perturbation configurations
and an ensemble of frequencies. 
Random matrix theory predicts that, after all non-universal effects
have been removed, the resulting scattering parameter, which we
denote as $s=(i\xi-1)/(i\xi+1)=|s|e^{i\varphi_{s}}$, should be
distributed uniformly in phase in $0\leq \varphi_{s}\leq 2\pi$, and
this result is independent of loss, frequency, and mean level
spacing \cite{Poisson_Kernel_Original, Henry_paper}. 
The previous work of Refs. \cite{S1, Sameer,Sirko} removed the
non-universal properties by performing normalization with only the
radiation impedance, as $i\xi=(Z-iX_{R})/R_{R}$.
Here we add ray trajectory corrections based on the maximum
trajectory length $L$,
\begin{equation}\label{eq:SOC_normalized_impedance}
    i\xi^{(L)}=(Z-iX_{avg}^{(L)})/R_{avg}^{(L)},
\end{equation}
and use the $\chi^{2}$ test to evaluate how uniform the resulting
phase distributions are. $R_{avg}^{(L)}$ and $X_{avg}^{(L)}$ are the
analytic continuations of the real and imaginary parts of
$Z_{avg}^{(L)}$ as $k\rightarrow k+ik/(2Q)$ in the experimental case
with loss. 
Experimental distributions of the phase $\varphi_{s}$ of $s$ are
calculated from 100 realizations and different frequency windows.
$\displaystyle\chi^{2}=\frac{1}{\langle n_{i}\rangle}
\sum_{i=1}^{N}(n_{i}-\langle n_{i}\rangle)^{2}$ measures the
deviation between the experimental distributions of $\varphi_{s}$
and a perfectly uniform distribution, where $n_{i}$ is the number of
elements in the $i^{th}$ bin in the histogram (with ten bins,
$N=10$) of the probability of the phase of the scattering parameter
$P(\varphi_{s})$, and $\langle n_{i} \rangle$ is the average of
$n_{i}$ over $i$. A smaller $\chi^{2}$ value means the experimental
data are closer to the theoretical prediction. 

Fig. \ref{chi_square} shows the averaged $\chi^{2}$ evaluated over
the spectral range from 6 to 18 GHz. The results indicate that the
distributions of the measured data are systematically more uniform
as more ray trajectories are taken into account in the impedance
normalization (Eq. (\ref{eq:SOC_normalized_impedance})).
The improvement is dramatic after including just a few short ray
trajectories ($L=50$ cm, 7 trajectories) and saturates beyond
$L=100$ cm (36 trajectories). The periodic wiggles represent the
effects of the strongest remaining trajectory not taken into account
in the
theory. 
Thus, we see that non-universal effects of ray trajectories in the
ensemble of wave-chaotic systems can be efficiently removed by
considering a few short ray trajectories or by increasing the window
size for the frequency ensemble.

\begin{figure}
\includegraphics[height=2.4in,width=3.0in]{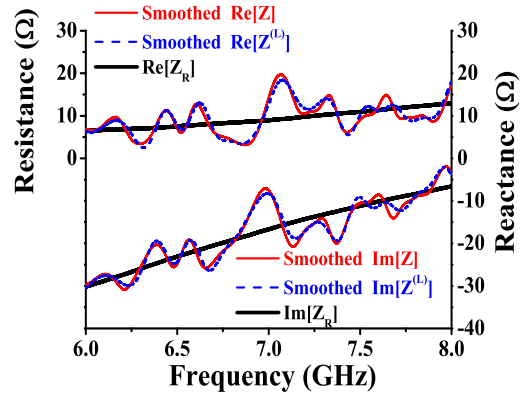}
\caption{(Color online) Plot of the smoothed impedance versus
frequency from 6 to 8 GHz. Shown are the real (three upper curves)
and the imaginary part (three lower curves) of the smoothed
impedance for the theory ($Z^{(L)}$ with $L=200$ cm, blue dashed)
and the experiment (red solid), as well as the measured radiation
impedance of the port ($Z_{R}$, black thick).}
\label{smoothed_impedance}
\end{figure}

In addition to experiments with ensemble averaging over perturber
positions, we now examine the theory in the stringent case of just a
single realization without scatterers, and we use only a frequency
ensemble.
We consider frequency smoothed experimental data and compare it with
the smoothed theoretical prediction. Fig. \ref{smoothed_impedance}
shows that the smoothed measured impedance $Z$ (red solid) agrees
with the smoothed theoretical impedance $Z^{(L)}$ (blue dashed).
Notice that $Z_{avg}^{(L)}\rightarrow Z^{(L)}$ because there is only
a single realization now. The smoothing function is a Gaussian with
standard deviation $\sigma=240$ MHz, which inserts an effective
low-pass Gaussian filter on the trajectory length, thus limiting the
terms in Eq. (\ref{eq:Z_avg_introduction}) to those with a path
length $l\lesssim c/\sigma=125$ cm. The results in Fig.
\ref{smoothed_impedance} shows that the theory correctly captures
systematic contributions from short trajectories. 

In conclusion, the non-universal effects of coupling and short ray
trajectories on wave scattering in chaotic systems are predicted by
a newly developed theory \cite{Hart} and verified experimentally.
This is accomplished through statistical tests of the scattering
parameter and comparisons of impedance in an ensemble of perturbed
systems, as well as a single-realization wave-chaotic system.
In particular, non-universal effects have been better considered and
removed from measured data to reveal underlying universally
fluctuations in the scattering parameter. 
These results should be useful in many fields, such as nuclear
scattering, atomic physics, quantum transport in condensed matter
systems, electromagnetics, acoustics, geophysics, etc.

We acknowledge seminal discussions with R. E. Prange
and assistance from Michael Johnson. 
This work was supported by the Air Force Office of Scientific
Research Grant No. FA95500710049.


\end{document}